# Characterisation of the Intel RealSense D415 Stereo Depth Camera for Motion-Corrected CT Perfusion Imaging

**Authors**

Mahdieh Dashtbani Moghari[1], Philip Noonan[2], David L. Henry[3,4], Roger R. Fulton[3,4,5], Noel Young[6,7], Krystal Moore[6], Andrew Evanns[3,8] and Andre Z. Kyme[1,4†]

**Affiliations**

1. School of Biomedical Engineering, Faculty of Engineering and Computer Science, University of Sydney, Sydney, Australia
2. School of Biomedical Engineering & Imaging Sciences, King's College London, UK
3. Sydney School of Medicine and Health, University of Sydney, Sydney, Australia
4. Brain and Mind Centre, University of Sydney, Sydney, Australia
5. Department of Medical Physics, Westmead Hospital, Sydney, Australia
6. Department of Radiology, Westmead Hospital, Sydney, Australia
7. Medical Imaging Group, School of Medicine, Western Sydney University, Sydney, Australia
8. Department of Aged Care & Stroke, Westmead Hospital, Sydney, Australia

† Corresponding author.

**Corresponding author details:**

E: <andre.kyme@sydney.edu.au>

**Abstract-** Despite using a protocol lasting only 1 min, head movement is problematic in cerebral CT perfusion (CTP) imaging of acute stroke, causing data inconsistencies that can compromise the quantitative haemodynamic modelling driving treatment decisions. Accordingly, retrospective motion correction is typically a default pre-processing step. Frame-to-frame registration is the most common form of retrospective correction but neglects the fact that motion is generally continuous, not discrete. By contrast, external tracking devices provide continuous motion monitoring and thereby the opportunity to fully correct the acquired data for motion, both between frames and within frames. The aim of this study was to characterise the Intel D415 stereo depth camera, a compact low-cost markerless tracking device, in terms of its suitability for retrospective CTP motion correction. The characterisation consisted of (i) thermal stability and noise jitter performance for a realistic static head phantom; (ii) validation of static and dynamic pose measurement accuracy against ground-truth robotic motion; and (iii) adaptation of the Intel D415 to a clinical CT scanner and validation of dynamic human head pose measurement accuracy against the marker-based OptiTrack system. The results showed that jitter was stable and thermally-induced pose drift was ≤ 1.5 mm and ≤ 0.5° during the first 10 - 20 min, after which it also became stable. For static poses, the mean difference between the Intel D415 estimates and commanded robot poses was ≤ 1.24 ± 0.01 mm and ≤ 0.68 ± 0.01° for position and orientation, respectively. For dynamic poses measured while the phantom travelled smooth continuous trajectories with median speed 0.031 $ms^{-1}$ (speed range 0 - 0.500 $ms^{-1}$), the root-mean-square-error (RMSE) was ≤ 1.40 ± 0.12 mm and ≤ 0.24 ± 0.02°. When tracking a simulated patient head trajectory derived from a clinical CTP scan, the average RMSE was ≤ 0.86 ± 0.03 mm and ≤ 0.16 ± 0.03°. Tracking the head motion of a human volunteer inside a clinical CT scanner, the average RMSE was ≤ 2.72 ± 0.24 mm and ≤ 0.55 ± 0.07°. Overall, our results suggest that a single D415 tracking system can achieve promising pose estimation accuracy, though slightly worse than typical CT resolution for brain scans, including CTP. This error is likely to be reduced to a practical level by combining multiple devices and this should be investigated in future work.

## 1. Introduction

Patient head movement during cerebral computed tomography perfusion (CTP) imaging of suspected acute stroke cases introduces image artifacts that can compromise lesion characterisation and the ensuing treatment decisions in at least 20% of CTP studies [1-3]. Compared to a standard neurological CT scan that completes within a fraction of a second, the likelihood of head motion is greater in CTP imaging due to the dynamic acquisition of 30-40 frames over a relatively long scanning time of ~1-2 minutes.

Head movement can cause CTP data inconsistencies in two different ways. The first is fast and discrete motion resulting in rigid misalignment between reconstructed frames, termed inter-frame motion corruption. The second is continuous motion resulting in data inconsistencies within individual frames, termed intra-frame motion corruption. Motion-corrupted frames may manifest visible artifacts like blurring, streaking and ghosting [4, 5], however the more serious impact is on the estimation of perfusion parameters derived from haemodynamic modelling [6]. These parameters, including cerebral blood volume (CBV), cerebral blood flow (CBF), mean transit time (MTT), and time-to-peak (TTP), are crucial for the lesion characterisation that drives treatment decision-making [7, 8]. Therefore, mitigating or correcting for patient head motion during imaging to improve the performance of image-based stroke analysis is highly desirable.

The use of restraining devices such as a foam headrest can limit patient head movement during image acquisition but does not completely prevent it [9]. More restrictive methods increase discomfort, reducing their acceptability by patients and potentially exacerbating movement [10]. An alternative approach to reduce the probability of motion artifacts is to use CT scanners with faster gantry rotation or more detector rows and/or X-ray sources in order to image the brain more rapidly [5, 11]. However, the timescale of the passage of contrast agent through the brain and the need to image the washout phase for modelling represents an important physiological constraint: although each frame can acquire faster using advanced hardware (reducing intra-frame motion), the total duration of the protocol – and thus the propensity for motion – remains unchanged.

Due to the limitations of motion prevention strategies, use of retrospective motion correction is a standard pre-processing step in the CTP image data analysis pipeline. By far the most common motion correction strategy is to rigidly register each individual CTP frame to a reference frame, for example the initial frame or another image such as a non-contrast CT [6, 12]. This data-driven frame-to-frame alignment can only correct inter-frame motion, not the intra-frame motion caused by continuous head movement. For continuous motion correction, external motion tracking may provide an advantage.

The rationale for external motion tracking in CTP imaging is two-fold: accuracy and continuity. Compared to data-driven techniques, external tracking can in principle deliver more accurate motion estimates because it is independent of the motion-corrupted data and not limited by the intrinsic resolution of the CT. Continuous tracking is also important because it is not reasonable to assume motion is purely inter-frame. In general, individual frames will also be corrupted by motion to some extent. Interestingly, all reports of motion-corrected CTP imaging focus on inter-frame motion, and the impact of intra-frame motion on the haemodynamic modelling of CTP data is not well understood.

Optical or infra-red (IR) motion tracking that relies on the detection of reflective discs or spheres attached to the head can provide highly accurate estimates of rigid-body head motion, typically at 10-100 Hz over the course of a scan. This is contingent on markers remaining rigidly coupled to the head [11, 13], which in practice is difficult to achieve using non-invasive attachment methods. Challenges related to the reliability and practicality of marker-based motion tracking systems have motivated a push to markerless systems [14, 15]. In a markerless approach, features used for object pose estimation are typically distinctive natural features or based on superimposed patterned (structured) illumination [15, 16]. Computer vision techniques can then be used for stereo reconstruction of sparse landmarks or point clouds from which changing object pose may be estimated [17]. Obtaining dense point clouds characterising the surface of an object at high frame rate has been made possible through a variety of consumer-grade devices such as the Microsoft Kinect [18-20] and Intel depth cameras [21-24].

Low cost and compact depth sensors are an attractive option for integration into medical imaging applications provided they meet the accuracy requirements. The Intel RealSense depth sensors were introduced onto the market in 2015 and were quickly deployed in a variety of motion tracking applications, including in healthcare [24-27]. The D400-series, introduced in 2018, feature hardware-accelerated stereo vision calculations and support the use of unstructured light. To date, these sensors have not been characterised for CT applications.

The aim of this study was to investigate the feasibility of applying the Intel RealSense D415 stereo depth camera for accurate and reliable motion estimation to enable continuous motion correction of CT/CTP studies. We characterise the performance of this device with respect to thermal stability and jitter, static and dynamic six degree-of-freedom (DoF) pose accuracy, and adaptability to the clinical setting.

## 2. Material and methods

### 2.1 Intel RealSense D415

The Intel D415 is a stereo-optical system consisting of right and left IR cameras (imagers) and an RGB sensor, enabling the capture of both depth and colour information in a scene, and an optional IR projector (Figure 1). The IR projector enhances depth estimation for low-textured objects by projecting a structured pattern onto the scene. The scene is captured by the imagers and the raw data sent to an onboard vision processor for depth map estimation. The depth map is generated via stereo matching, where the disparity (horizontal displacement) between a pair of corresponding pixels in the left and right images is calculated for each pixel. Thus, each pixel of the depth map represents the average distance to a region of the surface of the scene captured in the field-of-view (FoV).

The Intel D415 uses camera sensors consisting of 1280 × 720 pixels with pixel size 1.4 µm × 1.4 µm. The device weighs 72 g and has dimensions 99 × 20 × 23 mm (length × depth × height). Key technical specifications of the Intel D415 are summarised in Table 1.

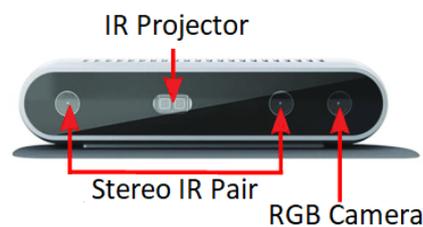

**Figure 1.** Intel RealSense D415 stereo depth camera (Intel 2021).

**Table 1.** Specifications of the Intel RealSense D415.

| Component | Specification |
|---|---|
| Horizontal field of view (FoV) | 65° ± 2° |
| Vertical FoV | 40° ± 1° |
| Diagonal FoV | 72° ± 2° |
| Depth map resolution | Up to 1280 × 720 pixels |
| Frame rate | Up to 90 frames per second (fps) |
| Minimum working distance | 0.16 m |
| Maximum range | ~10 m |
| Connector | USB Type-C |

## 2.2 Pose tracking

Our pose tracking software is an implementation of the signed distance function (SDF) tracker algorithm [28], which uses a GPU-parallelised iterative closest point approach to register sequential depth frames from the Intel RealSense onto a static volume. Depth frames are converted into a 3D point cloud where each point is defined by an $x$, $y$ and $z$ position and represents a location on the surface of the object being imaged. When the tracking software is initialised, the point cloud associated with the first frame is transformed into a SDF volume in which each voxel contains the distance to the nearest surface, implicitly describing the scene volumetrically. Subsequent depth frames are converted to point clouds before being registered to the SDF by sampling the SDF values at the $x$, $y$ and $z$ location of each point in the 3D GPU memory and iteratively optimising the rigid transform aligning the current frame to the SDF volume space.

With the depth camera fixed, any motion of the object in the tracking volume is measured as a relative transformation of the tracking volume with respect to the coordinate frame of the Intel depth camera (Figure 2a). A coordinate transformation can map this pose change to any arbitrary world frame. Real-time pose tracking rates (≥ 10 Hz) can be achieved on a wide range of GPU-enabled hardware. The pose tracking and estimation software was developed in C++ and OpenGL for cross compatibility between Windows and Linux based systems.

## 2.3 Intel D415 performance characterisation: benchtop

### 2.3.1 Experimental setup

The repeatability, stability and static and dynamic 6 DoF pose estimation accuracy of the Intel D415 was characterised using a head phantom manipulated by the 6 DoF UR3 robot (Universal Robots, Odense Denmark). The UR3 is highly flexible (± 360° rotation in all joints) and has pose repeatability of ± 0.1 mm and ± 0.01° [29, 30] and positional accuracy < 0.35 mm [31].

The experimental setup is shown in Figure 2. A realistic rubber human face mask phantom was rigidly mounted to the UR3 tool centre point (TCP) and viewed using a single Intel D415 rigidly fixed on a tripod at a nominal working distance of 0.25 m. The Intel D415 was connected to a powered USB3 hub interfacing with the host computer in order to stabilise the frame rate and avoid frame drop-out. Neither the Intel D415 IR projector or the RGB camera images were used in any of our experiments.

As the phantom was manipulated, the UR3 pose and associated timestamp were extracted in real time at 125 Hz from the robot controller MODBUS server via a dedicated port on the existing TCP/IP connection. At the same time, the Intel D415 recorded the pose of the phantom in its native coordinate frame (Figure 2a) at 30 fps. Default pre-sets were used for the Intel D415 [32] and the depth frame resolution was set to 1280 × 720 pixels.

For all of the robotic experiments we computed the median speed and speed range for the motion trajectories tested. The reported speed corresponded to the phantom nose-tip, a point 100 mm along the +*Z* direction of the phantom coordinate frame (Figure 2a). Speed of this point was computed from the MODBUS data for the TCP and using an offset transformation as described in [29].

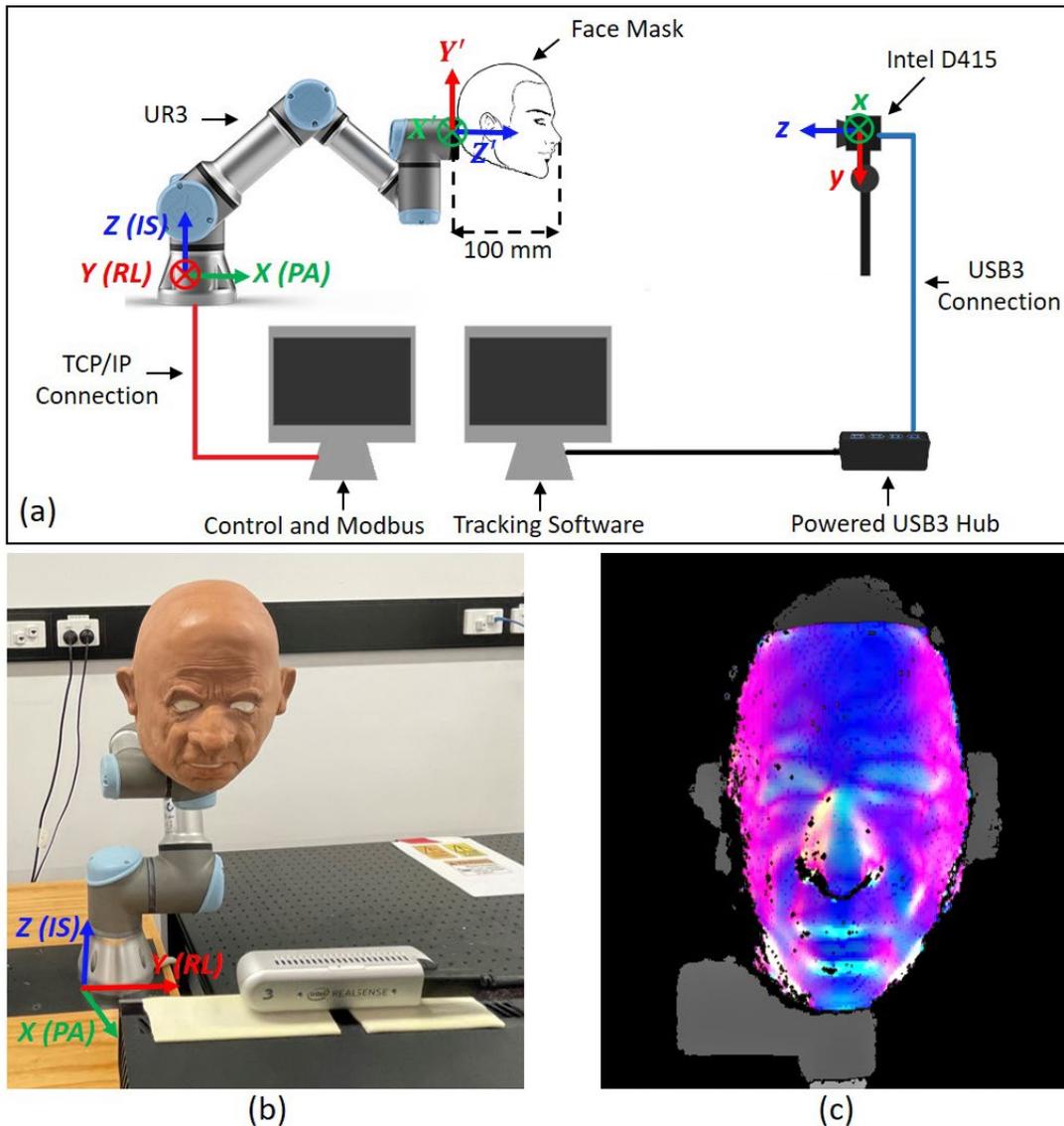

**Figure 2.** Benchtop setup for robotic motion tracking. (a) Schematic of the experimental setup to control the Intel D415 and robotic phantom. The moving phantom frame coincided with the default UR3 tool frame ($X'Y'Z'$). The robot base frame ($XYZ$) is defined based on posterior to anterior (PA), right to left (RL), and inferior to superior (IS). $xyz$ is the Intel D415 coordinate frame. (b) Rubber human face mask phantom mounted to the UR3 and viewed by the Intel D415. (c) Depth map generated by the Intel D415 using a region-of-interest confined to the phantom face.

### 2.3.2 System cross-calibration and synchronisation

To directly compare the commanded UR3 poses with the Intel D415 measurements, we required (i) a cross-calibration of the two systems so that the Intel D415 pose measurements could be reported in the robot frame, and (ii) precise temporal synchronisation of the pose measurements from the two devices.

To cross-calibrate the UR3 and Intel D415, 35 discrete poses of the phantom were measured simultaneously using the two devices. The calibration poses were chosen to sample all translational DoF (right–left ($RL$), inferior–superior ($IS$), and posterior-anterior ($PA$)) and rotational DoF ($rRL$, $rIS$, and $rPA$) (Figure 2a). At each calibration pose of the phantom, the Intel D415 and UR3 sampled the pose multiple times and pose averaging was used to obtain estimates with reduced noise and improved stability [33]. By default, the UR3 expresses three-dimensional (3D) rotations using axis-angle format, in which a 3-element vector describes the axis of rotation and its norm describes the magnitude of rotation. This format was converted to a rotation matrix and then combined with the position vector to represent the pose as a $4 \times 4$ homogeneous transformation matrix, $T$, according to:

$$T = \begin{bmatrix} \mathbf{R} & \mathbf{t}^T \\ 0 & 1 \end{bmatrix} \quad (1)$$

where $\mathbf{R}$ and $\mathbf{t}$ are a $3 \times 3$ rotation matrix and 3-element translation vector, respectively.

The cross-calibration matrix $\mathbf{X}$ was obtained using the hand-eye method in [34]. $\mathbf{X}$ relates pose changes in the robot and Intel D415 frames according to:

$$\Delta \mathbf{A}_{ij}^{r} = \mathbf{A}_i \, \mathbf{A}_j^{-1} \quad (2)$$

$$\Delta \mathbf{B}_{ij}^{I} = \mathbf{B}_i^{-1} \, \mathbf{B}_j \quad (3)$$

$$\Delta \mathbf{A}_{ij}^{r} \, \mathbf{X} = \mathbf{X} \, \Delta \mathbf{B}_{ij}^{I} \quad (4)$$

where $\mathbf{A}$ and $\mathbf{B}$ are homogeneous transformation matrices describing the absolute pose of an object in the UR3 ('r') and Intel D415 ('I') coordinate systems, respectively, and $\Delta \mathbf{A}_{ij}^{r}$ and $\Delta \mathbf{B}_{ij}^{I}$ are homogeneous transformation matrices describing a pose change (movement) between pose $i$ and $j$ in these respective coordinate systems [35]. In our experiment we considered pose changes with respect to the first pose, so $i = 1$ and $j \in \{1, 2, \ldots, n\}$, where $n$ is the total number of poses. According to equation (4), a pose change in the UR3 coordinate system is obtained according to:

$$\Delta \mathbf{A}_{ij}^{r} = \mathbf{X} \, \Delta \mathbf{B}_{ij}^{I} \, \mathbf{X}^{-1} \quad (5)$$

The synchronisation of the UR3 and Intel D415 pose measurements was achieved by manually aligning the respective motion traces, using the first transition of the phantom from stationary-to-moving and the last transition of the phantom from moving-to-stationary as reference points. Since the rate of pose measurement was different for the Intel D415 (30 fps) and UR3 MODBUS (125 Hz), the UR3 measurements were re-sampled to 30 fps before performing cross-calibration or pose comparisons.

### 2.3.3 Stability assessment

To assess the thermal dependency and noise stability (jitter) of the Intel D415, the head phantom (Figure 2) was tracked stationary for 70 minutes, repeated three times at the same pose. The amplitude and variance of the thermal drift and jitter were computed as a function of time from the raw Intel D415 pose estimates.

### 2.3.4 Static and dynamic pose accuracy

#### 2.3.4.1 Static performance

The robot was programmed to translate the phantom ±40 mm along each axis ($RL$, $IS$ and $PA$) and ±10° around each axis ($rRL$, $rIS$, and $rPA$), stopping for 3 s at each pose. The 12 discrete poses recorded by the Intel D415 were compared with the commanded UR3 poses after applying the cross-calibration and averaging over the samples collected during the 3 s per pose. The experiment was repeated five times to assess the variance.

### 2.3.4.2 Dynamic performance

The dynamic performance of the Intel D415 was assessed in three experiments: (1) tracking the 1D pose-pose transition ±40 mm along each axis ($RL$, $IS$ and $PA$) and ±10° around each axis ($rRL$, $rIS$, and $rPA$) at median speed of 0.262 ms$^{-1}$ (range 0 - 0.717 ms$^{-1}$); (2) tracking a continuous motion trajectory through 35 distinct arbitrary poses at three different median speeds: 0.031 ms$^{-1}$ (range 0 - 0.501 ms$^{-1}$), 0.022 ms$^{-1}$ (range 0 - 0.261 ms$^{-1}$), and 0.006 ms$^{-1}$ (range 0 - 0.061 ms$^{-1}$); and (3) tracking a simulated patient head motion sequence derived from a clinical CTP scan. This particular scan was identified by the reporting radiologist as having a degree of motion likely to impact clinical decision-making. The patient motion trace was obtained from this scan by rigidly registering each of the 33 reconstructed CTP frames to the first frame using SPM12 [36]. The $X$, $Y$, and $Z$ directions of the SPM12 coordinate system correspond to $RL$, $PA$ and $IS$, respectively. To simulate the patient motion track during the CTP acquisition, we matched the UR3 and SPM12 coordinate systems using a mapping transform and programmed the UR3 to visit the 33 mapped poses within 60 s. This corresponded to a median speed of 0.020 ms$^{-1}$ (range 0 - 0.093 ms$^{-1}$).

For all three experiments, poses recorded by the Intel D415 were compared to the UR3 poses after applying the cross-calibration. The root-mean-square-error (RMSE) and mean difference for each DoF were used to assess the pose accuracy. All experiments were repeated five times and the average RMSE or difference was reported.

## 2.4 Intel D415 performance characterisation: clinical environment

To evaluate the performance of the Intel D415 in a clinical setting, the device was adapted to a clinical CT scanner to track both phantom and human head motion. The Intel D415 was mounted to the gantry of a Siemens Biograph mCT scanner using a custom 3D-printed attachment facilitating pan/tilt positioning and connected to a powered USB3 hub interfacing with the host computer. Performance was measured relative to the OptiTrack IR motion tracking system (NaturalPoint, Inc.). Three synchronised OptiTrack cameras (1280 × 1024 resolution, 100 fps) were clamped to the wall at the rear of the CT scanner. All cameras had direct line-of-sight to the head inside the gantry at a nominal working distance of 2 m. The OptiTrack multi-camera configuration was calibrated using the standard calibration wand procedure before collecting measurements. The experimental setup is shown in Figure 3.

### 2.4.1 System cross-calibration and synchronisation

To directly compare the Intel D415 pose estimates against the OptiTrack, the two systems were cross-calibrated using the hand-eye method as described in Section 2.3.2. The rubber face mask was positioned on the CT bed facing the Intel D415 and centrally located in its FoV. Five IR reflective markers were affixed to the scalp of the phantom for tracking using the Optitrack, and all three OptiTrack cameras had line-of-sight to every marker to maximise tracking accuracy. The phantom was moved manually to 35 discrete calibration poses within the FoV of both tracking systems at 5-10 s intervals and simultaneously tracked by both devices. Pose averaging [33] was used to obtain estimates with reduced noise before computing **X** in equation (4).

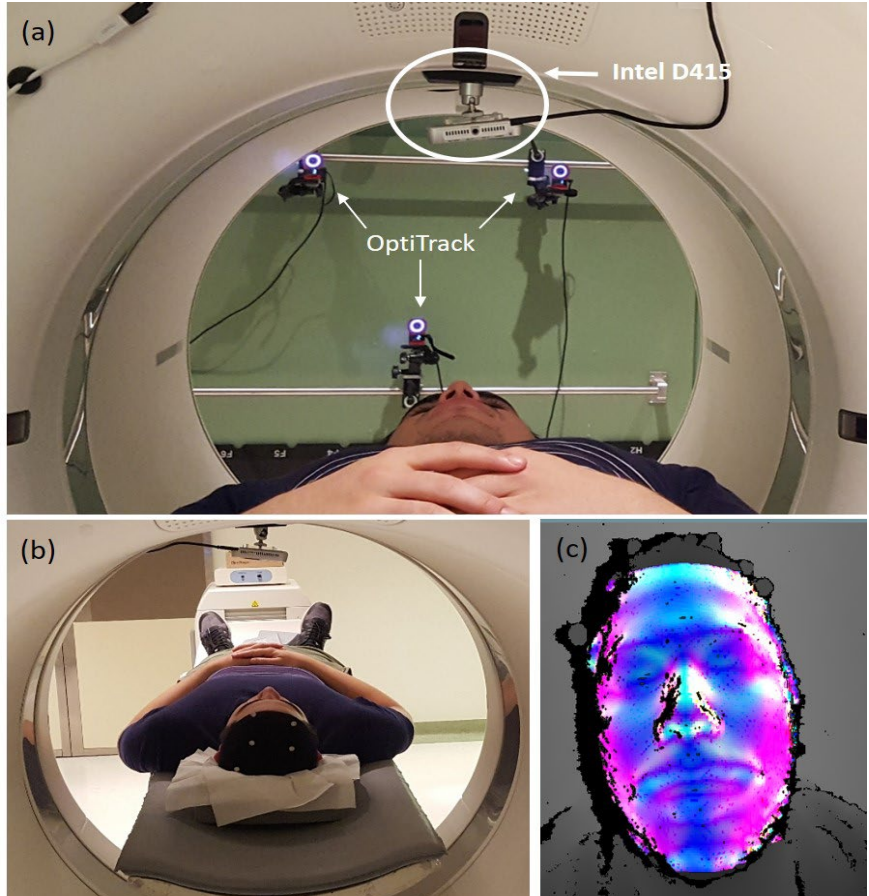

**Figure 3.** Evaluating the Intel D415 performance in a clinical CT scanner. (a) Experimental setup showing head tracking of a volunteer using the mounted Intel D415 (white circle) and simultaneously using the IR-based OptiTrack system with 3 cameras; (b) IR-reflective markers affixed to the OptiTrack head cap (shown here for the human volunteer, but the same markers were applied for the head phantom test); (c) depth map generated by the Intel D415 using a region-of-interest confined to the human volunteer face.

### 2.4.2 Phantom experiments

The head phantom was positioned inside the CT scanner and moved continuously by hand in three separate 60-s trials while being tracked by the two systems. The Intel D415 pose measurements were compared quantitatively with the OptiTrack estimates after applying the cross-calibration.

### 2.4.3 Human volunteer experiments

A human volunteer wore a tightly fitted fabric head cap affixed with five IR reflective markers defining a rigid-body registered with the OptiTrack system (Figure 3b). The volunteer lay in the scanner and performed three 60-s head motion sequences while their head motion was tracked by the OptiTrack and Intel D415 systems simultaneously. The volunteer was instructed to perform slow and continuous arbitrary head motion in each of the three trials. The amplitude of the motion was consistent with 'mild-to-considerable' motion as defined in [1] and involved all DoF. The Intel D415 head motion trajectories were again compared against the OptiTrack estimates after applying the cross-calibration.

## 3. Results

### 3.1 Intel D415 performance characterisation: benchtop

#### 3.1.1 Stability assessment

The Intel D415 tracking of a static phantom over 70 minutes (Figure 4) showed a gradual drift in the pose measurement as the sensor internal temperature increased from start-up. The pose estimate stabilised after 10-20 minutes, once the internal temperature of the device reached a steady state. The average internal temperature change from start-up to stabilisation over three trials was 12.66 ± 2.08 °C. The absolute pose drift during this time was ≤ 1.5 mm and ≤ 0.5° with respect to translational and rotational DoFs, respectively. The largest jitter amplitude, computed as the maximum amplitude range, was 0.5 mm ± 0.1 mm for translational measurements and was observed for the $RL$ and $IS$ axes. The maximum jitter amplitude for rotations was approximately 0.1° for all three axes.

#### 3.1.2 Static pose accuracy performance

Table 2 shows the static localisation accuracy of the Intel D415, reported as the difference between the commanded UR3 pose and mean Intel D415 pose for six translational and six rotational poses, averaged over five trials. The accuracy was ≤ 1.24 ± 0.01 mm and ≤ 0.68 ± 0.01° for translation and rotation, respectively.

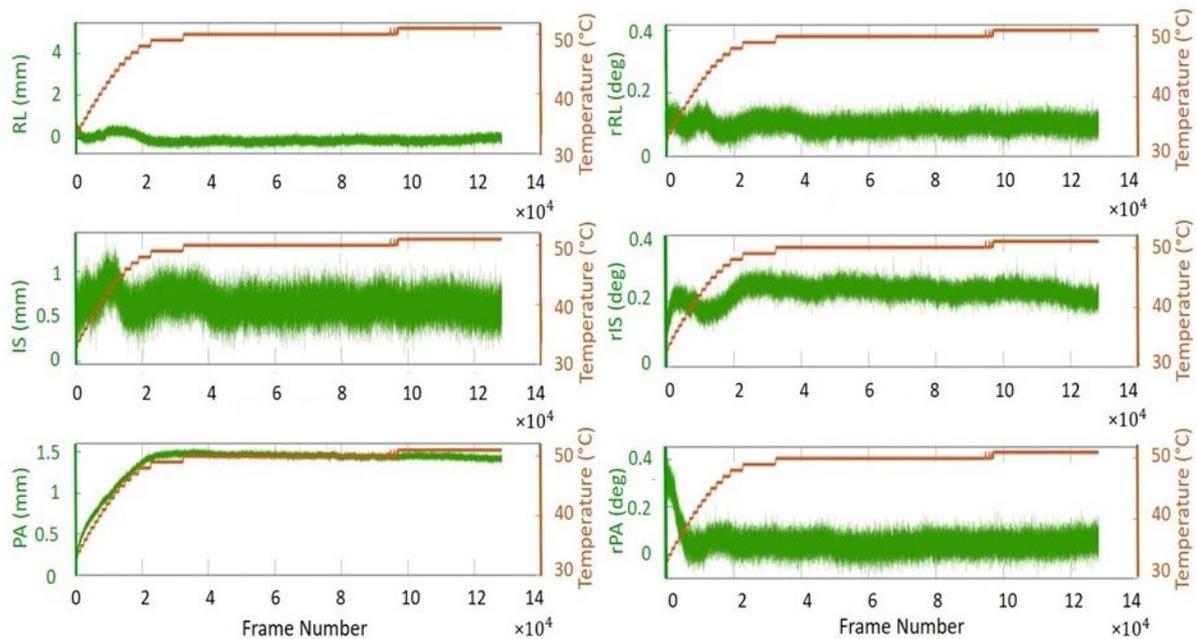

**Figure 4.** Representative thermal stability data for the Intel D415. Pose measurements from start-up (green) and the corresponding internal temperature change (orange) over 70 min (approximately 130,000 frames).

**Table 2.** 1-D static and dynamic pose accuracy of the Intel D415 compared to the UR3 robot. Measurements are reported as mean ± SD.

| Direction | Commanded UR3 position | Static accuracy[a] (mean difference) | Dynamic accuracy[b] (mean RMSE) | Dynamic accuracy[b] (mean difference) |
|---|---|---|---|---|
| $RL$ (mm) | 40 | 1.24 ± 0.01 | 1.08 ± 0.63 | 0.98 ± 0.74 |
|  | -40 | 0.56 ± 0.07 | 0.78 ± 0.31 | 0.06 ± 0.79 |
| $IS$ (mm) | 40 | -0.71 ± 0.02 | 1.26 ± 0.70 | -1.05 ± 0.65 |
|  | -40 | 0.23 ± 0.03 | 0.77 ± 0.39 | 0.29 ± 0.63 |
| $PA$ (mm) | 40 | 0.67 ± 0.05 | 0.60 ± 0.09 | 0.50 ± 0.08 |
|  | -40 | 0.05 ± 0.04 | 0.4 ± 0.13 | 0.32 ± 0.17 |
| $rRL$ (°) | 10 | -0.10 ± 0.01 | 0.05 ± 0.01 | -0.02 ± 0.01 |
|  | -10 | 0.01 ± 0.01 | 0.07 ± 0.01 | -0.06 ± 0.01 |
| $rIS$ (°) | 10 | 0.68 ± 0.01 | 0.37 ± 0.04 | 0.25 ± 0.07 |
|  | -10 | -0.02 ± 0.01 | 0.67 ± 0.09 | 0.57 ± 0.08 |
| $rPA$ (°) | 10 | -0.05 ± 0.01 | 0.04 ± 0.01 | -0.02 ± 0.02 |
|  | -10 | 0.01 ± 0.01 | 0.04 ± 0.01 | -0.03 ± 0.02 |

(a) Static localisation accuracy is expressed as the difference between the commanded UR3 position (ground-truth) and average pose reported by the Intel D415.
(b) Dynamic accuracy is expressed as either the RMSE or difference between commanded and measured poses, averaged across all trials.

**Table 3.** Dynamic pose accuracy of the Intel D415 compared to the UR3 robot.

| Direction | Metric | Arbitrary motion $(S_1)$[a] | Arbitrary motion $(S_2)$[a] | Arbitrary motion $(S_3)$[a] | Patient motion[b] |
|---|---|---|---|---|---|
| $RL$ (mm) | Mean RMSE ± SD | 1.40 ± 0.12 | 1.38 ± 0.02 | 1.25 ± 0.01 | 0.53 ± 0.04 |
| $IS$ (mm) |  | 1.13 ± 0.27 | 0.91 ± 0.04 | 0.85 ± 0.02 | 0.86 ± 0.03 |
| $PA$ (mm) |  | 0.55 ± 0.12 | 0.50 ± 0.04 | 0.45 ± 0.04 | 0.77 ± 0.07 |
| $rRL$ (°) |  | 0.12 ± 0.02 | 0.10 ± 0.00 | 0.09 ± 0.00 | 0.11 ± 0.01 |
| $rIS$ (°) |  | 0.22 ± 0.01 | 0.22 ± 0.01 | 0.20 ± 0.00 | 0.16 ± 0.03 |
| $rPA$ (°) |  | 0.24 ± 0.02 | 0.23 ± 0.00 | 0.14 ± 0.00 | 0.03 ± 0.01 |
| $RL$ (mm) | Mean difference ± SD | -0.24 ± 0.01 | 0.23 ± 0.01 | 0.14 ± 0.01 | 0.40 ± 0.05 |
| $IS$ (mm) |  | -0.11 ± 0.02 | 0.10 ± 0.01 | 0.10 ± 0.01 | 0.19 ± 0.01 |
| $PA$ (mm) |  | -0.38 ± 0.02 | -0.4 ± 0.04 | -0.35 ± 0.03 | 0.49 ± 0.01 |
| $rRL$ (°) |  | -0.04 ± 0.00 | 0.01 ± 0.00 | 0.01 ± 0.00 | -0.09 ± 0.01 |
| $rIS$ (°) |  | 0.06 ± 0.00 | -0.01 ± 0.00 | -0.01 ± 0.00 | 0.07 ± 0.01 |
| $rPA$ (°) |  | 0.13 ± 0.00 | 0.14 ± 0.00 | 0.06 ± 0.00 | -0.01 ± 0.00 |

(a) The arbitrary continuous trajectory travelled at a different median speed of $S_1$ = 0.031 ms$^{-1}$ (range 0 - 0.501 ms$^{-1}$), $S_2$ = 0.022 ms$^{-1}$ (range 0 - 0.261 ms$^{-1}$) and $S_3$ = 0.006 ms$^{-1}$ (range 0 - 0.061 ms$^{-1}$).
(b) Simulated head motion trajectory of a patient during a 60 s CTP scan, median speed 0.020 ms$^{-1}$ (range 0 - 0.093 ms$^{-1}$).

### 3.1.3 Dynamic pose accuracy performance

Table 2 also shows the dynamic accuracy and repeatability of the Intel D415 measurements relative to the UR3 for 1D transitions from the origin to six translational and six rotational end positions. The average RMSE was ≤ 1.26 ± 0.70 mm and ≤ 0.67 ± 0.09° for translation and rotation, respectively, and the corresponding absolute mean difference was ≤ 1.05 ± 0.65 mm and ≤ 0.57 ± 0.08°.

Table 3 shows the dynamic localisation accuracy of the Intel D415 measurements relative to the UR3 for the arbitrary continuous trajectory consisting of 35 distinct poses, and for three different speed profiles. The mean RMSE for translation and rotation was ≤ 1.40 ± 0.12 mm and ≤ 0.24 ± 0.02° for median speed 0.031 ms$^{-1}$ (range 0 - 0.501 ms$^{-1}$), ≤ 1.38 ± 0.02 mm and ≤ 0.23 ± 0.00° for median speed 0.022 ms$^{-1}$ (range 0 - 0.261 ms$^{-1}$), and ≤ 1.25 ± 0.01 mm and ≤ 0.20 ± 0.00° for median speed 0.006 ms$^{-1}$ (range 0 - 0.061 ms$^{-1}$). Figure 5 shows the Intel D415 motion estimates against the UR3 poses for the arbitrary dynamic path traversed at a median speed of 0.022 ms$^{-1}$ (range 0 - 0.261 ms$^{-1}$).

The dynamic localisation accuracy of the Intel D415 relative to the UR3 for simulated patient head movement is shown in Table 3 (right column) and Figure 6. The mean RMSE was ≤ 0.86 ± 0.03 mm and ≤ 0.16 ± 0.03° and absolute mean difference was ≤ 0.49 ± 0.01 mm and ≤ 0.09 ± 0.01.

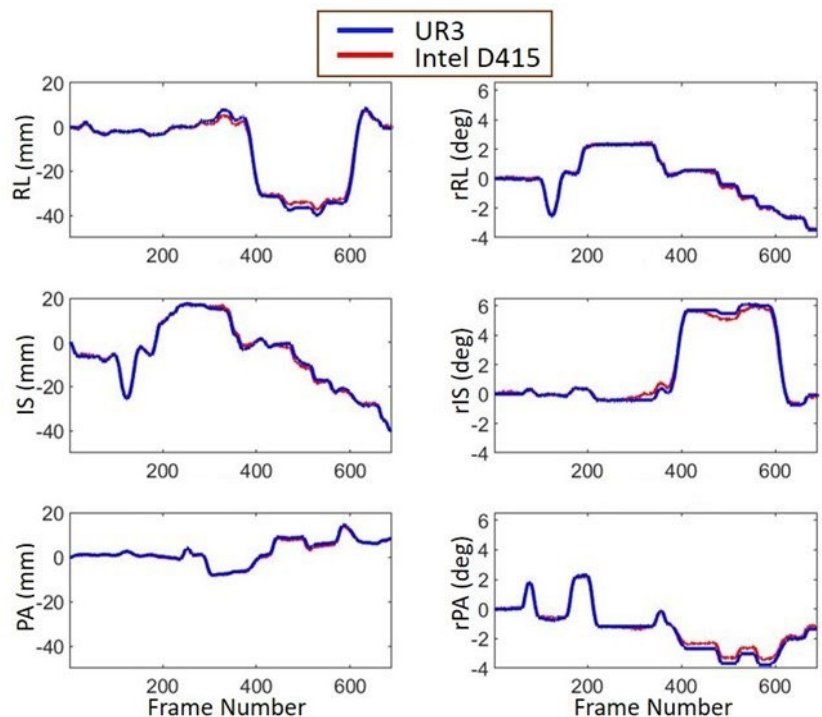

**Figure 5.** Comparison of motion estimates for the Intel D415 (red) and UR3 (blue) for an arbitrary continuous motion trajectory traversed at a median speed of 0.022 ms$^{-1}$ (range 0 - 0.261 ms$^{-1}$).

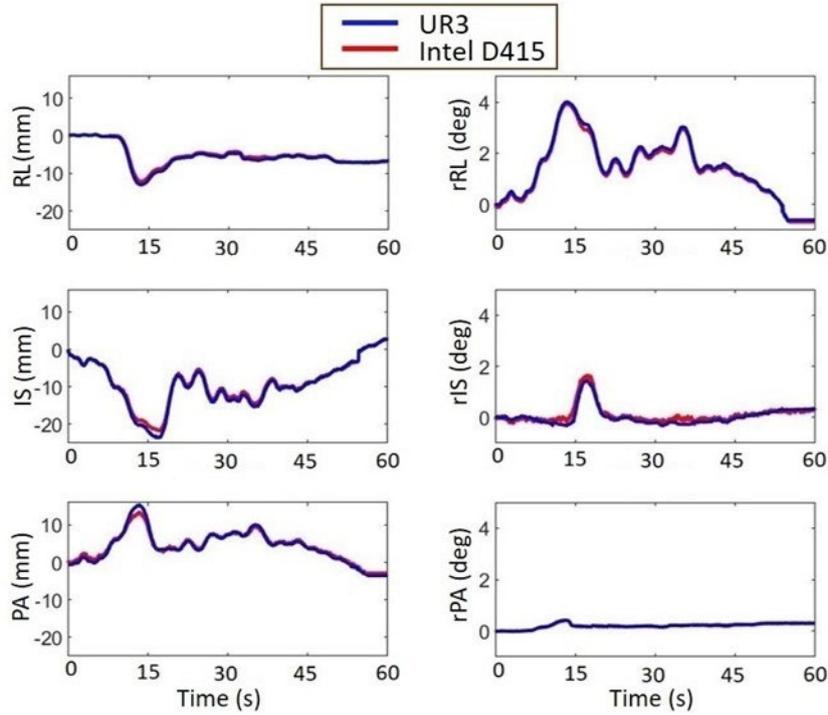

**Figure 6.** Comparison of motion estimates for the Intel D415 (red) and UR3 (blue) for a simulated patient head motion trajectory during a 60 s CTP scan.

**Table 4**. Dynamic pose accuracy of the Intel D415 relative to the OptiTrack for the phantom and volunteer motion inside a clinical CT scanner.

| Motion | Metric | RL (mm) | IS (mm) | PA (mm) | $rRL$ (°) | $rIS$ (°) | $rPA$ (°) |
|---|---|---|---|---|---|---|---|
| Phantom | Mean RMSE | 3.74 | 3.53 | 2.24 | 0.91 | 0.76 | 0.68 |
|  | ± SD | ± 2.39 | ± 2.99 | ± 1.34 | ± 0.66 | ± 0.59 | ± 0.47 |
|  | Mean difference | 1.09 | 1.13 | -0.66 | -0.19 | 0.05 | -0.26 |
|  | ± SD | ± 3.16 | ± 3.77 | ± 2.58 | ± 1.53 | ± 1.11 | ± 0.82 |
| Volunteer | Mean RMSE | 2.72 | 1.70 | 0.81 | 0.32 | 0.55 | 0.47 |
|  | ± SD | ± 0.24 | ± 0.41 | ± 0.14 | ± 0.08 | ± 0.07 | ± 0.40 |
|  | Mean difference | 0.34 | -1.12 | -0.36 | -0.17 | -0.02 | -0.06 |
|  | ± SD | ± 0.47 | ± 2.34 | ± 0.67 | ± 0.26 | ± 0.06 | ± 0.53 |

### 3.2 Intel D415 performance characterisation: clinical environment

Table 4 shows the dynamic localisation accuracy of the Intel D415 relative to the OptiTrack for the phantom and volunteer experiments in a clinical setting. For the head phantom, the mean RMSE for translation and rotation was ≤ 3.74 ± 2.39 mm and ≤ 0.91 ± 0.66°, and the absolute mean difference was ≤ 1.13 ± 3.77 mm and ≤ 0.26 ± 0.82°. Pose measurements from each device are compared in Figure 7 for each DoF.

For volunteer head movement, the mean RMSE for translation and rotation was ≤ 2.72 ± 0.24 mm and ≤ 0.55 ± 0.07°, and the absolute mean difference was ≤ 1.12 ± 2.34 mm and ≤ 0.17 ± 0.26°. Pose measurements from each device are compared in Figure 8 for each DoF.

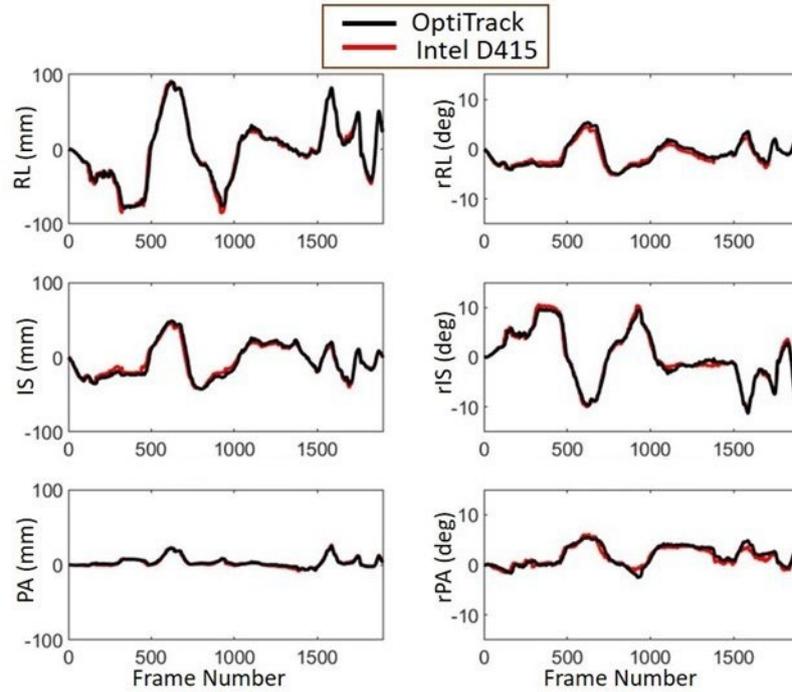

**Figure 7.** Comparison of motion estimates for the Intel D415 (red) and OptiTrack (black) when tracking a head phantom inside a clinical CT scanner for 60 s.

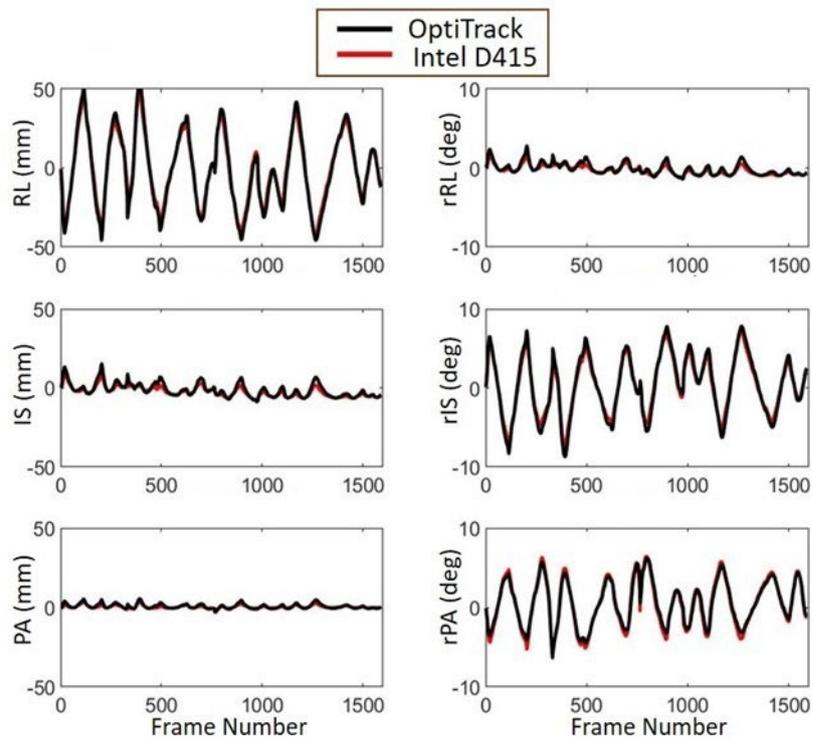

**Figure 8.** Comparison of motion estimates for the Intel D415 (red) and OptiTrack (black) when tracking human head movement for 60 s inside a clinical CT scanner.

## 4. Discussion

This study reports a characterisation of the Intel D415 depth sensor performance to evaluate its potential use in head motion tracking for motion-corrected CT/CTP imaging. Thermal stability and measurement jitter were assessed over 70-min trials and static and dynamic pose estimation accuracy was validated against ground-truth robotic motion in benchtop experiments and against a multi-camera stereo-optical tracking system in the clinical environment.

Although thermally-induced pose drift was observed (Figure 4), the impact appears to be negligible over the 1-min scanning time in CTP imaging. Nevertheless, the accuracy of pose measurements cannot be assumed until thermal stability has been reached, approximately 10-20 min post power-up. This is also when cross-calibration should be performed to ensure that it is consistent with the steady state of the device.

Measurement jitter amplitude remained stable throughout 70-min trials and was also unaffected by changes in the Intel D415 internal temperature. Jitter is likely due to noise in the stereo matching and uncertainty in the iterative process used to calculate the position and orientation of the object. Although we showed that the jitter is stable, the jitter amplitude may be object-dependent. This is a potential limitation that was not investigated in our study. In the benchtop evaluation involving robotic motion, the static pose accuracy of the Intel D415 appeared to be DoF-dependent. The poorest accuracy was evident for shifts along the $RL$ axis (mean difference 1.24 ± 0.01 mm) and for head roll about the $IS$ axis (mean difference 0.68 ± 0.01°) axis. Motion related to these DoFs is more likely to result in occlusion near the nose (Figure 2, 3). Occlusion occurs when pixels cannot be seen by both the left and right imagers, precluding stereo reconstruction of that region which ultimately manifests as a gap in the point cloud model. Since the nose is a region of high topological change on the face, it helps to achieve robust depth-map alignment. Accordingly, gaps in this area may reduce the reliability of pose estimation. A possible solution to reduce the likelihood of occlusion and thereby improve pose estimation accuracy is to use multiple depth sensors to track the object from multiple views. This will be investigated in future work.

As shown in Table 3, the Intel D415 showed sub-millimetre accuracy in tracking the simulated patient motion trace (mean RMSE ≤ 0.86 ± 0.03 mm and ≤ 0.16 ± 0.03°). However, a conservative accuracy estimate (mean RMSE) based on all experiments involving the robotically controlled mask (Table 2 and 3) was < 2 mm and < 1° for translations and rotations, respectively. What is also evident from Table 3 is a trend towards improved accuracy as the speed of motion reduces. This is very typical of camera-based motion tracking systems and is usually due to the reduced motion blur in the raw data at slower object speed.

Compared to the robotic motion testing, the dynamic accuracy of the Intel D415 was poorer when validating it against the OptiTrack inside the CT scanner (Table 4). Possible reasons for this include: (i) less control over motion speed and range compared to robotic motion, (ii) increased likelihood of object occlusion and of the object leaving the FoV, and (iii) potential pose estimation error in the OptiTrack (ground-truth) due to deformation of the head cap resulting in non-rigid motion. The latter is a well-known challenge of marker-based tracking systems applied to human subjects [37].

Typically, CTP volumes comprise slices with 5 mm thickness and 0.43 mm in-plane resolution. Our results showed that the pose estimation accuracy of the Intel D415 was poorer than this in-plane image resolution, which implies that although the Intel D415 will aid motion correction, there will also be residual error. Moreover, since the pose accuracy of the Intel D415 appeared to be DoF-dependent, an optimised multi-sensor setup to track the object from multiple views could potentially reduce the impact of object position on pose estimation accuracy by improving the reliability of stereo matching. This will be considered in our future work.

High sample-rate ('continuous') motion tracking provides the opportunity to fully correct the acquired data for both inter- and intra-frame movement. A promising data-driven approach for CT has been demonstrated [38], however the practicality of this method for clinical application is still being optimised. Nevertheless, even with data-driven motion tracking and correction, there is value in complementary tracking which can provide robust pose initialisation and augment data gaps during tracking drop-out or when data-driven estimation is unreliable. This is where markerless tracking based on a device such as the Intel D415 may prove versatile, convenient and cheap compared to previous systems.

## 5. Conclusion

This study characterised the Intel D415 RealSense depth sensor as a markerless motion tracking device for continuous head pose estimation at close range (≤ 0.25 m), for potential use in motion-corrected CTP imaging. A conservative pose accuracy estimate for this device was < 2 mm and < 1° in bench-top experiments involving realistic head phantom and head motion, and < 3 mm and < 1° tracking a human face inside a clinical CT scanner. The Intel D415 therefore is a potential candidate as a cheap and versatile motion tracking device in a standalone capacity or to augment potential data-driven motion estimation techniques.